\documentclass[prl, twocolumn, superscriptaddress, nofootinbib]{revtex4-1}

\usepackage[english]{babel}
\usepackage{hyperref}

\usepackage[pdftex]{graphicx}
\usepackage{color}
\usepackage[toc,page]{appendix}

\usepackage{amsmath}
\usepackage{amssymb}
\usepackage{mathtools}
\usepackage{braket}

\usepackage[sort&compress]{natbib}

\newcommand{\un}[1]{\ensuremath{\,\mathrm{#1}}}
\renewcommand{\v}[1]{\ensuremath{\boldsymbol{#1}}}
\newcommand{\dg}{\dagger}
\newcommand{\fig}[1]{Figure~\ref{fig:#1}}
\newcommand{\Fig}[1]{Figure~\ref{fig:#1}}

\newcommand{\eq}[1]{equation~(\ref{eq:#1})}
\newcommand{\lr}[1]{\ensuremath{\left( #1 \right)}}

\renewcommand{\Im}[1]{\ensuremath{\mathrm{Im} \left(#1\right)}}
\newcommand{\I}{\mathrm{i}}
\renewcommand{\ap}{\alpha}

\newcommand{\Sg}{\Sigma}

\newcommand{\Tr}[1]{\ensuremath{\mathrm{Tr}\left(#1\right)}}

\newcommand{\eps}{\varepsilon}
\newcommand{\bt}{\beta}
\newcommand{\dl}{\delta}

\begin{document}

\title{Current vortices in aromatic carbon molecules}

\author{Thomas~Stegmann}
\email{stegmann@icf.unam.mx}
\affiliation{Instituto de Ciencias F\'isicas, Universidad Nacional Aut\'onoma de M\'exico, 62210
  Cuernavaca, Mexico}

\author{John~A.~Franco-Villafa\~ne}
\affiliation{CONACYT - Instituto de F\'isica, Universidad Aut\'onoma de San Luis Potos\'i, 78290 San
  Luis Potos\'i, Mexico}

\author{Yenni~P.~Ortiz}
\affiliation{Instituto de F\'isica, Universidad Nacional Aut\'onoma de M\'exico, 04510 Mexico City, Mexico}

\author{Michael~Deffner}
\affiliation{Institute for Inorganic and Applied Chemistry, University of Hamburg, 20146 Hamburg,
  Germany}

\author{Carmen~Herrmann}
\affiliation{Institute for Inorganic and Applied Chemistry, University of Hamburg, 20146 Hamburg,
  Germany}

\author{Ulrich~Kuhl}
\affiliation{Universit\'{e} C\^{o}te d'Azur, CNRS, Institut de Physique de Nice, 06100 Nice, France}

\author{Fabrice~Mortessagne}
\affiliation{Universit\'{e} C\^{o}te d'Azur, CNRS, Institut de Physique de Nice, 06100 Nice, France}

\author{Francois~Leyvraz}
\affiliation{Instituto de Ciencias F\'isicas, Universidad Nacional Aut\'onoma de M\'exico, 62210
  Cuernavaca, Mexico}
\affiliation{Centro Internacional de Ciencias, 62210 Cuernavaca, Mexico}

\author{Thomas~H.~Seligman}
\affiliation{Instituto de Ciencias F\'isicas, Universidad Nacional Aut\'onoma de M\'exico, 62210
  Cuernavaca, Mexico}
\affiliation{Centro Internacional de Ciencias, 62210 Cuernavaca, Mexico}

\date{\today}

\begin{abstract}
  The local current flow through three small aromatic carbon molecules, namely benzene, naphthalene
  and anthracene, is studied. Applying density functional theory and the non-equilibrium Green's
  function method for transport, we demonstrate that pronounced current vortices exist at certain
  electron energies for these molecules. The intensity of these circular currents, which appear not
  only at the anti-resonances of the transmission but also in vicinity of its maxima, can exceed the
  total current flowing through the molecular junction and generate considerable magnetic
  fields. The $\pi$ electron system of the molecular junctions is emulated experimentally by a
  network of macroscopic microwave resonators. The local current flows in these experiments confirm
  the existence of current vortices as a robust property of ring structures. The circular currents
  can be understood in terms of a simple nearest-neighbor tight-binding H\"uckel model. Current
  vortices are caused by the interplay of the complex eigenstates of the open system which have
  energies close-by the considered electron energy. Degeneracies, as observed in benzene and
  anthracene, can thus generate strong circular currents, but also non-degenerate systems like
  naphthalene exhibit current vortices. Small imperfections and perturbations can couple otherwise
  uncoupled states and induce circular currents.
\end{abstract}

\maketitle

\section{Introduction}

In the last decades, the structural size of electronic circuits has been reduced enormously. This is
manifested impressively by the fact that the first field-effect transistor, invented by W.~Schockley
et al. in 1947, had the size of the palm of a hand, while nowadays, billions of transistors of size
of a few nanometers are packed on a single chip. This miniaturization of electronic circuits gave
rise to the question if even a single molecule can be used as the functional element of an
electronic device \cite{Aviram1974} and founded eventually the active research field of molecular
electronics, see for example Refs.~\cite{Nitzan2003, Cuniberti2005, Cuevas2017} for an overview. One
of the main research topics of molecular electronics is to understand the current flow through a
molecular junction. A recent milestone has been the observation of quantum interference in molecular
junctions \cite{Guedon2012, Manrique2015, Carlotti2016, Li2019} and its theoretical understanding
\cite{Solomon2008, Markussen2010, Berritta2015, Sangtarash2016, Yang2016, Nozaki2017b}. However, the
\textit{local current flow}, i.e. the spatial distribution of the current in the molecule, has
received less attention, see Refs.~\cite{Xue2004, Ernzerhof2006, Sai2007, Solomon2010,
  Herrmann2010b, Rai2010, Rai2012, Maiti2013, Bahamon2015, Al-Dirini2016, Lazzeretti2016,
  Monaco2016, Sundholm2016, Dimitrova2017, Nozaki2017, Li2017, Hansen2017, Stuyver2017, Fias2017,
  Jhan2017, Stuyver2018, Pohl2018, Cabra2018, Garner2018, Jensen2019, Garner2019, Patra2017,
  Dhakal2019, Taninaka2019} for recent work. This can be attributed to the fact that -- to the best
of our knowledge -- the local current flow in molecular junctions has been measured only indirectly
by nuclear magnetic resonance \cite{Troisi2007, Troisi2007b} but not yet directly.

One remarkable theoretical prediction with respect to the local current flow is the fact that
aromatic molecules can show at certain electron energies pronounced current vortices, which can
exceed even the total current flowing through the molecular junction. These predictions have been
made by means of state-of-the-art theory, combining Kohn-Sham density functional theory (DFT) for
the electronic structure of the molecular junction with the non-equilibrium Green's function method
(NEGF) for the electron transport \cite{Solomon2010, Herrmann2010b, Cabra2018}. It has also been
shown that the current vortices persist even if the molecular junction is described by a more
elementary theory, namely a tight-binding H\"uckel Hamiltonian, where electron-electron interactions
are not taken into account, and commonly only a single atomic orbital per atom is used
\cite{Ernzerhof2006, Rai2010, Rai2012, Maiti2013, Patra2017, Dhakal2019}. Also a mixture of both
approaches, the density functional based tight-binding (DFTB) method, predicts vortices in the local
current flow \cite{Nozaki2017}. Thermoelectric driven ring currents have also been reported recently
within a tight-binding model \cite{Rix2019}. Therefore, the observed circular currents are not a
subtle many-body phenomenon but a more robust general property of ring structures.

\begin{figure}[t]
  \centering
  \includegraphics[scale=0.3]{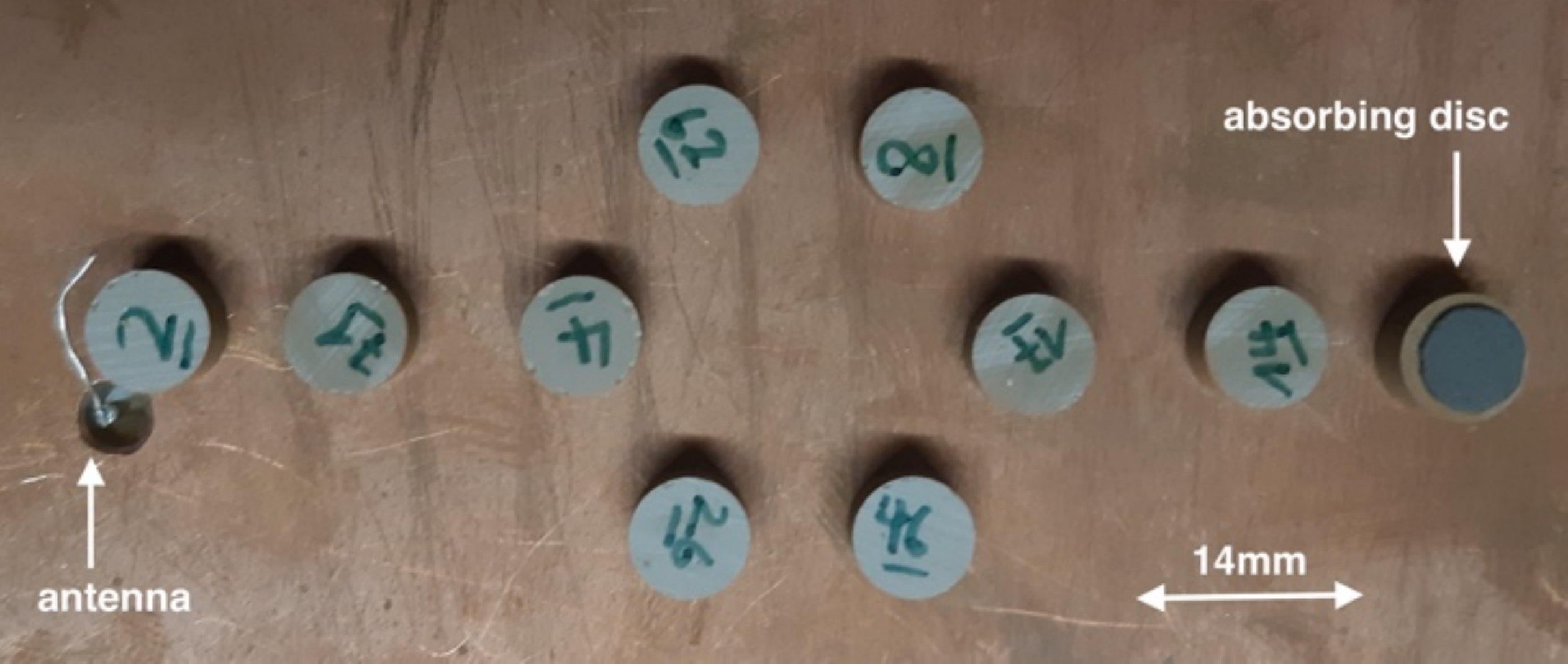}
  \caption{Photo of the microwave emulation experiment for the benzene molecule in the para
    configuration of the leads. Dielectric cylindrical resonators model the carbon atoms. The source
    antenna, which injects microwaves, is shown at the left end. A resonator with an absorbing
    material glued to its top models the drain at the right end. The electromagnetic field (or wave
    function) is measured on each resonator by a movable antenna through the metallic top plate
    (both not shown in the photo), which allows to calculate the local transmissions.}
  \label{fig:2}
\end{figure}

In this article, we analyze the current vortices in three small aromatic carbon molecules, namely
benzene, naphthalene and anthracene molecules, see \fig{1}. In particular, we show that circular
currents can not only be observed close to anti-resonances in the transmission function but also in
the vicinity of transmission maxima and hence, generate a magnetic field. Indeed, in the following,
we shall use a definition of circular current, see \cite{Rai2010, Dhakal2019}, which corresponds
precisely to the source of this magnetic field. More important, here we give experimental
indications of current vortices in ring structures analogous to the molecular structures. The formal
equivalence of the Helmholtz equation for electromagnetic waves and the non-interacting
Schr\"odinger equation for quantum wave functions allows to emulate quantum systems in microwave
cavities \cite{Stoeckmann1999, Kuhl2005, Laurent2007}. This idea can be used to realize
tight-binding systems by networks of dielectric microwave resonators \cite{Franco-Villafane2013,
  Poli2015}, as models for graphene \cite{Kuhl2010, Bellec2013, Bellec2013b, Boehm2015,
  Stegmann2017b} and molecules \cite{Stegmann2017}. Here we apply this method to emulate the current
flow in aromatic carbon molecules by means of a network of dielectric resonators, see \fig{2}. By
measuring the amplitude and phase of the electromagnetic field on each microwave resonator, we can
investigate experimentally the local current flow in these structures. The emulation experiments
show clearly circular currents as observed in the quantum calculations.

\section{Systems \& Methods}

\begin{figure}[t]
  \centering
  \includegraphics[scale=0.35]{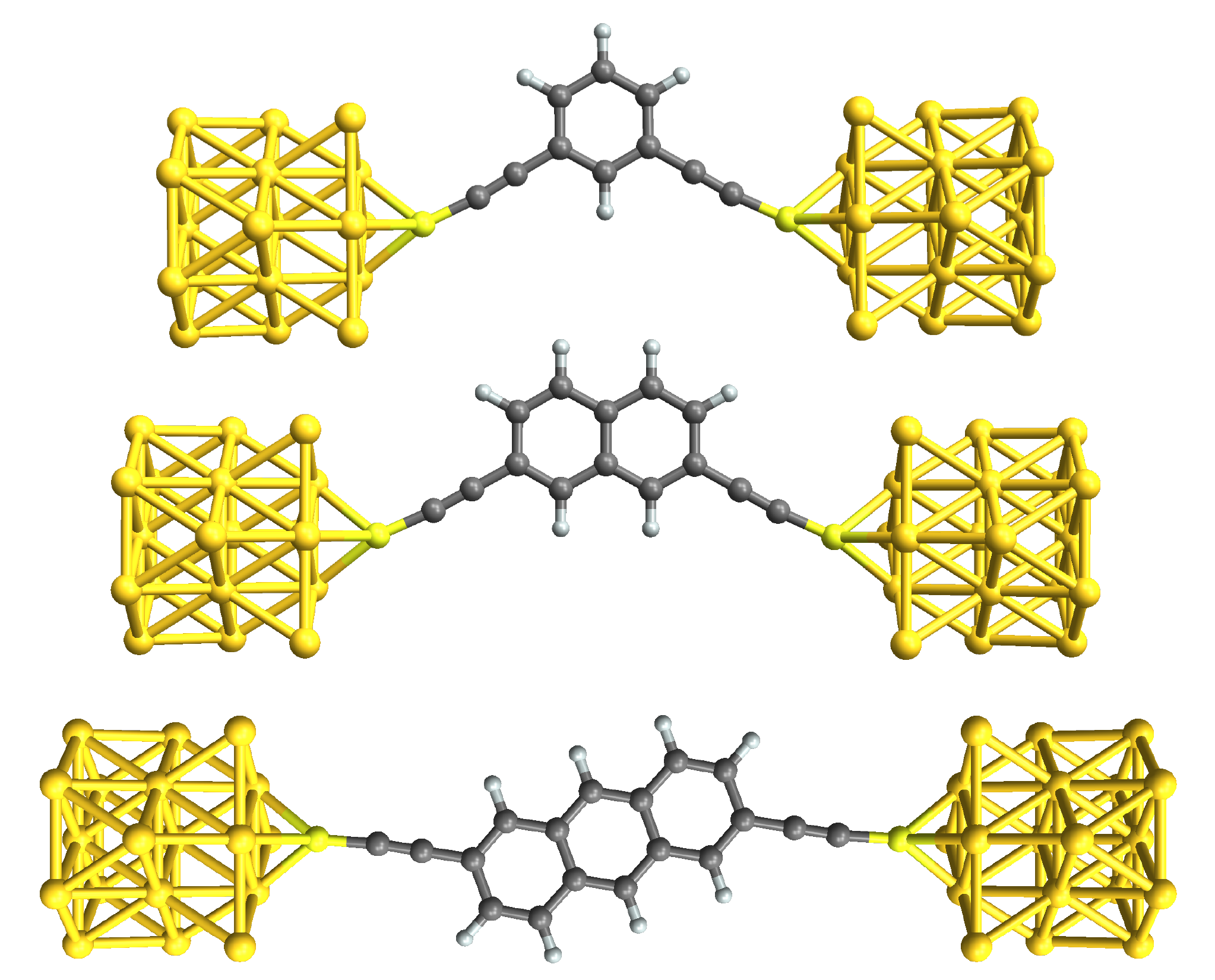}
  \caption{Benzene (top), naphthalene (middle) and anthracene (bottom) molecules are connected via
    two carbon spacer atoms and a sulfur atom to gold clusters (19 atoms, fcc lattice symmetry) and
    hence, form a molecular junction. Benzene and naphthalene are shown with leads in the meta
    position, while for anthracene the leads are in the para position. In our studies, both lead
    configurations are taken into account for all molecules.}
  \label{fig:1}
\end{figure}

We study the current flow in aromatic carbon molecules, to be precise, benzene, naphthalene and
anthracene. These molecules are attached to gold clusters using two additional carbon spacer atoms
and a sulfur atom, see \fig{1}. The gold clusters are built up of 19 atoms in fcc lattice symmetry,
though the precise configuration of the gold clusters does not change qualitatively the current flow
in the molecular junction.

\subsection{Quantum transport calculations}

In order to calculate the current flow through the molecular junction, the structure is optimized
first by means of DFT as implemented in the Gaussian09 software package \cite{Gaussian09}. The
exchange-correlation functional B3LYP and the basis set LANL2MB are used to describe the system. On
this basis, the current flow is calculated by means of the NEGF method as implemented in the Artaios
software package \cite{Artaios}. It calculates the Green's function of the scattering region
(i.e. the molecular junction without the gold clusters)
\begin{equation}
  \label{eq:1}
  G(E)= \lr{E\,S-H-\Sg_S-\Sg_D}^{-1},
\end{equation}
using the effective single-particle Kohn-Sham Hamiltonian $H$ and the overlap matrix $S$ from the
DFT optimization. The self-energies $\Sg_{S/D}$ describe the effect of the source and drain
reservoirs, which are attached through the gold clusters in order to inject and extract
electrons. These reservoirs are modelled within the wide-band approximation
\cite{Verzijl2013}. Using the Green's function, we study the overall transport by means of the
transmission function
\begin{equation}
  \label{eq:2}
  T(E)= 4 \Tr{\Im{\Sg_S}G\,\Im{\Sg_D}G^\dg},
\end{equation}
as well as the local transmissions between the atoms $i$ and $j$
\begin{equation}
  \label{eq:3}
  T_{ij}(E)= \Im{H_{ij}^*G^n_{ij}},
\end{equation}
where the correlation function is given by
\begin{equation}
  \label{eq:4}
  G^n= G \, \Im{\Sg_S} G^\dg.
\end{equation}
Note that in \eq{3} we have assumed implicitly the sum over pairs of local basis functions which
belong to the two atoms. In order to obtain the current from the transmission and the local
transmissions, one has to integrate these functions within the bias voltage window. Details of the
methods can be found in Refs.~\cite{Datta1997, Pecchia2004, Cuevas2017, Herrmann2010}.

\subsection{Microwave emulation experiments}

Non-interacting quantum systems can be emulated by means of a network of dielectric microwave
resonators \cite{Franco-Villafane2013, Poli2015, Kuhl2010, Bellec2013, Bellec2013b, Boehm2015,
  Stegmann2017b, Stegmann2017}. Here, we use this technique to gain insight into the local current
flow in aromatic carbon molecules. In the experiment, the carbon atoms are represented by identical
cylindrical resonators ($5\un{mm}$ height, $4\un{mm}$ radius, refractive index $n \approx 6$), that
are placed between two metal plates, see \fig{2}. The nearest neighbor distance of the resonators is
on average $14 \un{mm}$, while their precise distance ratios reflect the interatomic distances found
by the DFT optimization of the molecule. The coupling strength between neighboring resonators decays
approximately exponentially, as shown previously \cite{Stegmann2017, Bellec2013b}. Such an
exponential decay is a realistic model for slightly varying carbon-carbon bond distances, as shown
for deformed graphene \cite{Ribeiro2009}. The hydrogen atoms are not taken into account explicitly,
as we assume that the current is carried predominantly by the $\pi$~electron system that is formed
by the $2p_z$~orbitals of the carbon atoms. Hence, these orbitals, as well as their interactions,
are emulated by a microwave resonator network.

The effects of the gold clusters and sulfur atoms are taken into account indirectly by the antenna
at the left end that injects microwaves and a resonator with an absorbing material glued to its top
at the right end which serves as the drain. The microwaves are injected as transverse electrical
modes around the frequency $\nu_0= 6.65 \un{GHz}$, where the individual resonators have an isolated
resonance. Note that in contrast to the theory, the source (injecting antenna) and drain (absorbing
resonator) are modelled differently in the emulation experiment, compare \fig{1} and \fig{2}. The
other components of the molecular junction (gold, sulfur and hydrogen atoms), their complicated
interactions as well as correlations between the electrons are taken into account (approximately) in
the DFT-NEGF calculations but not in the emulation experiments. To which extent molecules that are
composed of different types of atoms can be modelled by means of different types of microwave
resonators will be addressed in our future work.

In order to determine the local transmission in the resonator network, the electromagnetic field (or
wave function) $\psi(\v{r})$ is measured on each resonator by means of a movable antenna through the
metallic top plate. Starting from the definition of the probability current density
$\v{j}(\v{r})= \frac{\hbar}{m} \Im{\psi^*(\v{r}) \nabla \psi(\v{r})}$ and discretizing it, we obtain
for the transmission between two resonators at frequency $\nu$ \cite{Seba1999}
\begin{equation}
  \label{eq:5}
  T^{\text{mw}}_{ij}(\nu)= \Im{d_{ij}^{-1} \, \psi_i^* \psi_j},
\end{equation}
where $d_{ij}$ is the distance between the resonators $i$ and $j$. Note that in the case of a
non-interacting system, \eq{5} is formally equivalent to \eq{3}, because
$G^{n}_{ij} \sim \psi_i^*\psi_j$ and $d_{ij} $ is proportional to the coupling matrix elements
$H_{ij}$.

%\vspace{3mm}

\begin{figure}[t]
  \centering
  \includegraphics[scale=0.75]{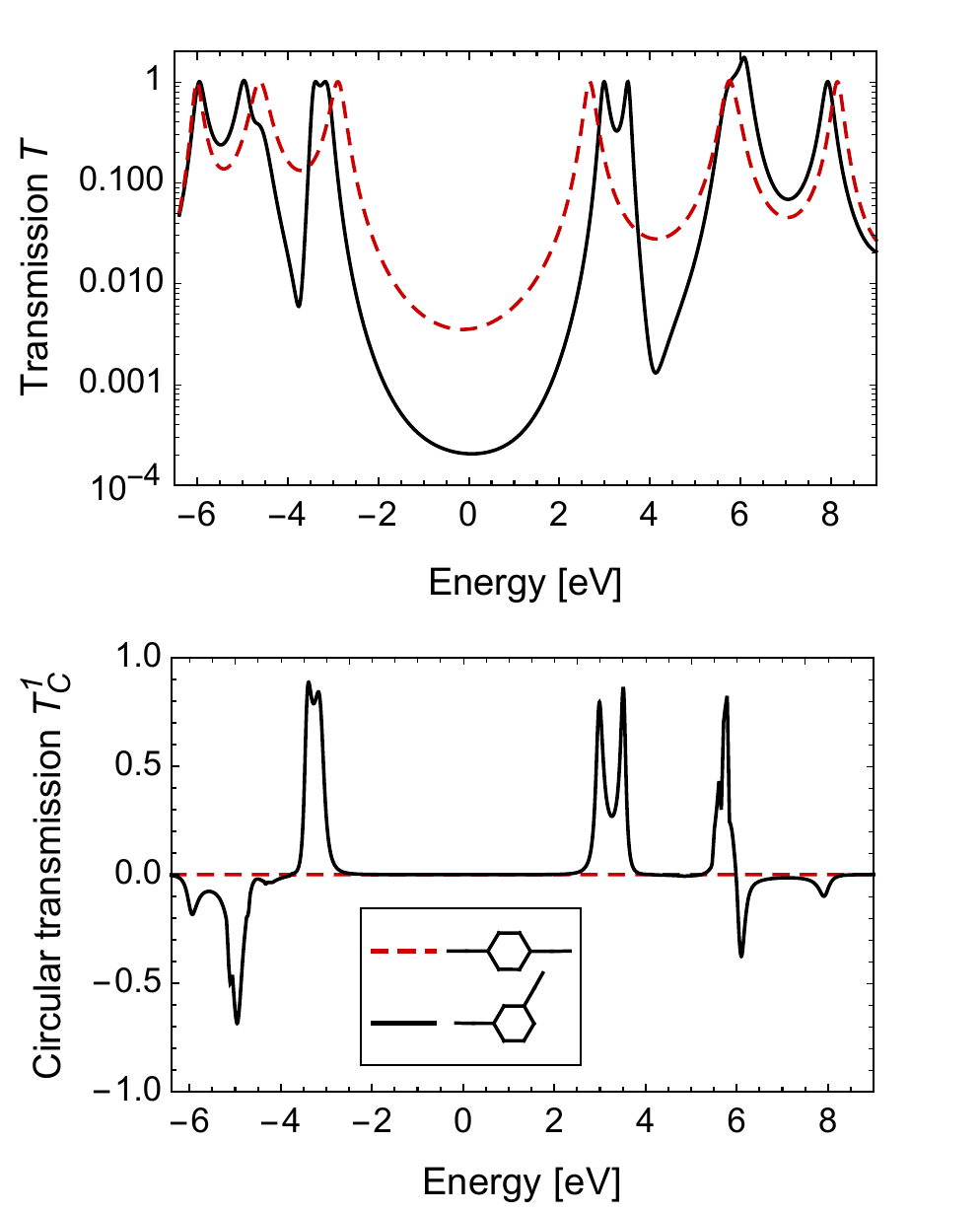}
  \caption{Calculated transmission $T$ (top) and circular transmission $T_C^1$ (bottom, calculated
    by \eq{6}) in the benzene molecule with leads in the meta (black curve) and the para (red dashed
    curve) configuration. The transmission shows in both cases band gaps, resonances as well as
    anti-resonances. Note that the middle between the HOMO and LUMO resonance energies has been
    chosen as the origin of the energy scale. Strong circular currents can be observed for certain
    electron energies in the meta configuration, while in the para configuration no current vortices
    exist.}
  \label{fig:3}
\end{figure}

\section{Results \& Discussion}
\label{sec:Results}

\subsection{Benzene}

% DFT-NEGF CALCULATIONS

\begin{figure}[htb]
  \centering
  \includegraphics[scale=0.42]{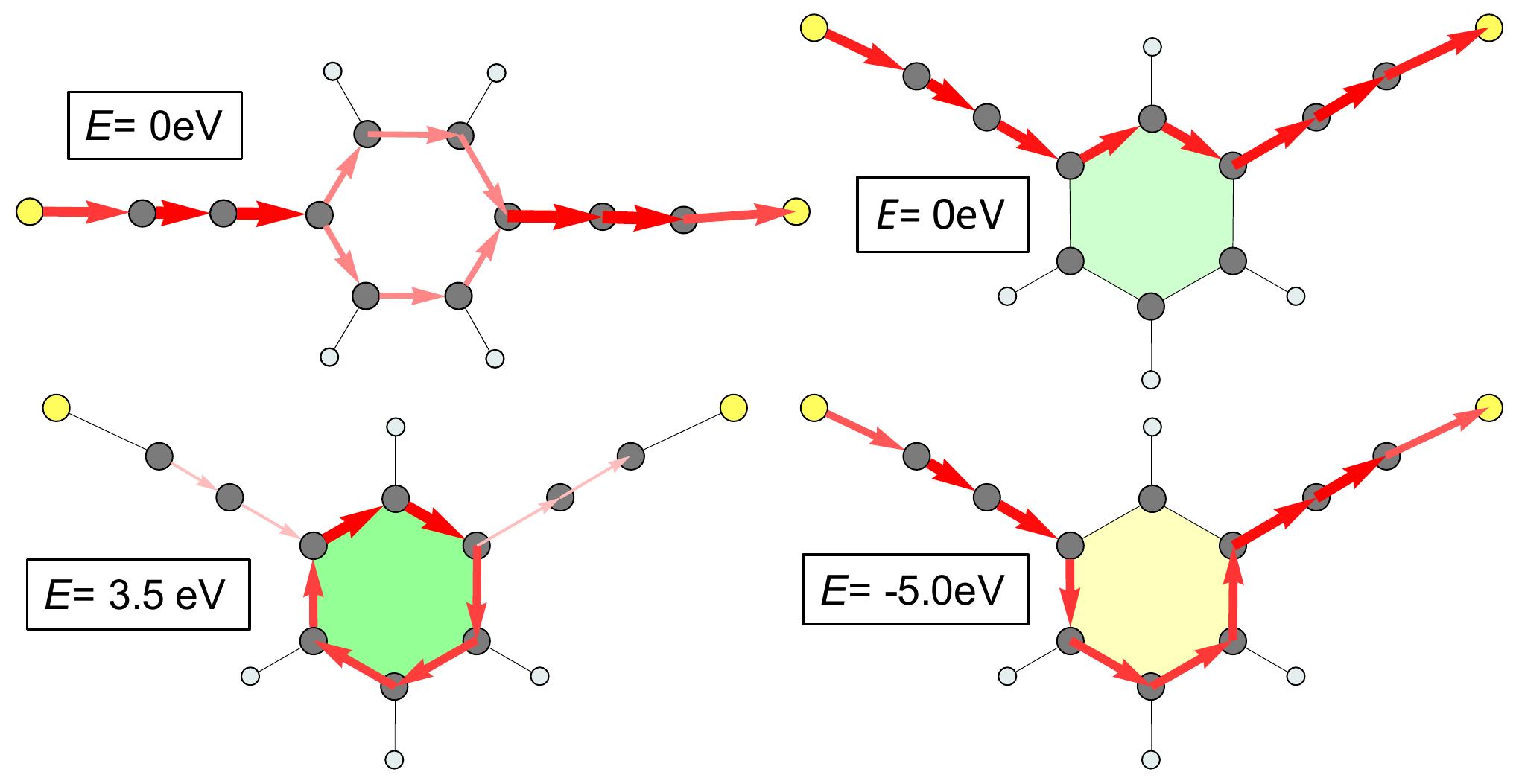}
  \caption{Normalized transmission pathways (arrows) in the benzene molecular junction, calculated
    by the DFT-NEGF method. The carbon, hydrogen and sulfur atoms are represented by dark gray,
    light gray and yellow disks, respectively. The gold clusters (see \fig{1}) are not shown. The
    size and color shading of the arrows indicate the magnitude of the local transmissions. The
    color shading of the hexagons indicates the normalized circular transmissions $T_C^1$ (see
    \eq{6}); greenish colors for clockwise circulation, yellowish colors for the opposite
    direction. For the above shown transmission pathways the full dependence on the electron energy
    is shown in the movie
    \href{https://www.fis.unam.mx/~stegmann/current-vortices/Benzene-DFT.mov}{\texttt{Benzene-DFT.mov}}}
  \label{fig:4}
\end{figure}

Let us begin our discussion with the simplest aromatic carbon molecule, namely benzene. The
transmission calculated by means of the DFT-NEGF method is shown in \fig{3} (top) for the para as
well as for the meta configuration of the leads. In both configurations, transmission band gaps,
resonances as well as anti-resonances can be observed. Instead of the global transport properties of
the molecular junction, which have been addressed previously \cite{Cuevas2017}, we rather
investigate here on which pathways the current flows through the system. Typical transmission
pathways, which can be observed by varying the electron energy, are shown in \fig{4} by means of
arrows of different size and color shading. Note that these local transmissions are normalized with
respect to the maximum value at each electron energy (or microwave frequency) and hence, do not
reflect the absolute transmission through the molecule. Moreover, transmissions less than 10\% of
the maximal value are not plotted.

Starting with the para configuration, we observe that the normalized transmission pathways are
independent of the electron energy and split up into two paths that carry equal amounts of
current. These properties can be understood easily by the mirror symmetry of the molecular junction
with respect to the horizontal axis through the center of the carbon hexagon, which implicates
identical currents in both paths and hence, circular currents are not possible.

In the meta configuration, we observe that within the transmission band gap ($E \sim 0 \un{eV}$),
the current is flowing on the shortest path between source and drain. Such asymmetric transmission
pathways are possible as the molecular junction is no longer mirror symmetric with respect to the
horizontal axis. If we change the electron energy to the transmission resonances, we observe a
circular current that is rotating around the carbon ring. Passing through the antiresonance (at
$E \sim 4 \un{eV}$), the direction of rotation changes. The full dependence of the transmission
pathways on the electron energy (or microwave frequency) is shown in the three movies
\href{https://www.fis.unam.mx/~stegmann/current-vortices/Benzene-DFT.mov}{\texttt{Benzene-DFT.mov}}
for the DFT-NEGF calculations,
\href{https://www.fis.unam.mx/~stegmann/current-vortices/Benzene-DFT.mov}{\texttt{Benzene-Experiment.mov}}
for the microwave emulations experiment, and
\href{https://www.fis.unam.mx/~stegmann/current-vortices/Benzene-DFT.mov}{\texttt{Benzene-TB.mov}}
for calculations within a simple tight-binding model. These movies can be accessed by clicking on
the file names.

The circular transmission pathways in the carbon hexagon $\ap$ are quantified by \cite{Rai2010,
  Dhakal2019}
\begin{equation}
  \label{eq:6}
  T_C^\ap(E)= \frac{1}{L_\ap} \, \sum_{\substack{i,j \in \ap \\[0.5mm] i<j}}
  T_{ij}(E) \, \text{sign}\lr{(\v r_i {-}\v r_\ap) \times (\v r_j {-}\v r_\ap)},
\end{equation}
where $T_{ij}(E)$ is the local transmission between two carbon atoms located at the positions
$\v r_i$ and $\v r_j$. The sum is over all atoms which are part of the hexagon $\ap$ with center
$\v r_\ap$ and length $L_\ap$.The circular transmission as a function of the electron energy is
shown in \fig{3} (bottom). The circular transmission, calculated in terms of the normalized
transmissions, is indicated in \fig{4} as the color shading of the carbon hexagons. Both figures
confirm the existence of circular currents in meta-benzene for certain energies ranges, whereas no
current vortices exist in para-benzene.

The circular transmissions give rise to a magnetic field passing through the carbon rings. In order
to estimate its strength, we integrate $T_C^1(E)$ around the double resonance at $ 3 \un{eV}$ within
a bias window of $2 \un{eV}$ (to be precise, in the energy range from $2\un{eV}$ to $4\un{eV}$) and
obtain an absolute circular current of approximately $I_C \sim 30 \un{\mu A}$, which in turn causes
in the center of the carbon hexagon a magnetic field%
\footnote{The magnetic field on an axis, which goes perpendicular through the center of a carbon
  hexagon, is given by
  \begin{equation*}
    B(z)= \frac{3 \sqrt{3} \mu_0 I_C}{4 \pi a} \frac{1}{\bigl[\lr{z/a}^2+3/4\bigr]\sqrt{\lr{z/a}^2+1}},
  \end{equation*}
  where $a \sim 0.139 \un{nm}$ is the distance of the carbon atoms and $I_C$ is the current
  circulating in the hexagon.}%
of $B \sim 150 \un{mT}$. Note that the circular transmission in \eq{6} is defined in such a way that
it is the only source of the magnetic field that is passing through the ring. In other words, the
remaining transverse current does not generate any mangetic flux, see Ref.~\cite{Rai2010,
  Dhakal2019} for details. Transmission pathways like those at $E=0 \un{eV}$ and $E=-5 \un{eV}$ are
also considered as current vortices, because they have a finite circular transmission and give rise
to a magnetic field. However, this magnetic field will be weaker and more inhomogenous than in the
cases where the current is circulating uni-directionally around the complete carbon ring as at
$E= 3.5 \un{eV}$.

% EXPERIMENT

\begin{figure}[t]
  \centering
  \includegraphics[scale=0.52]{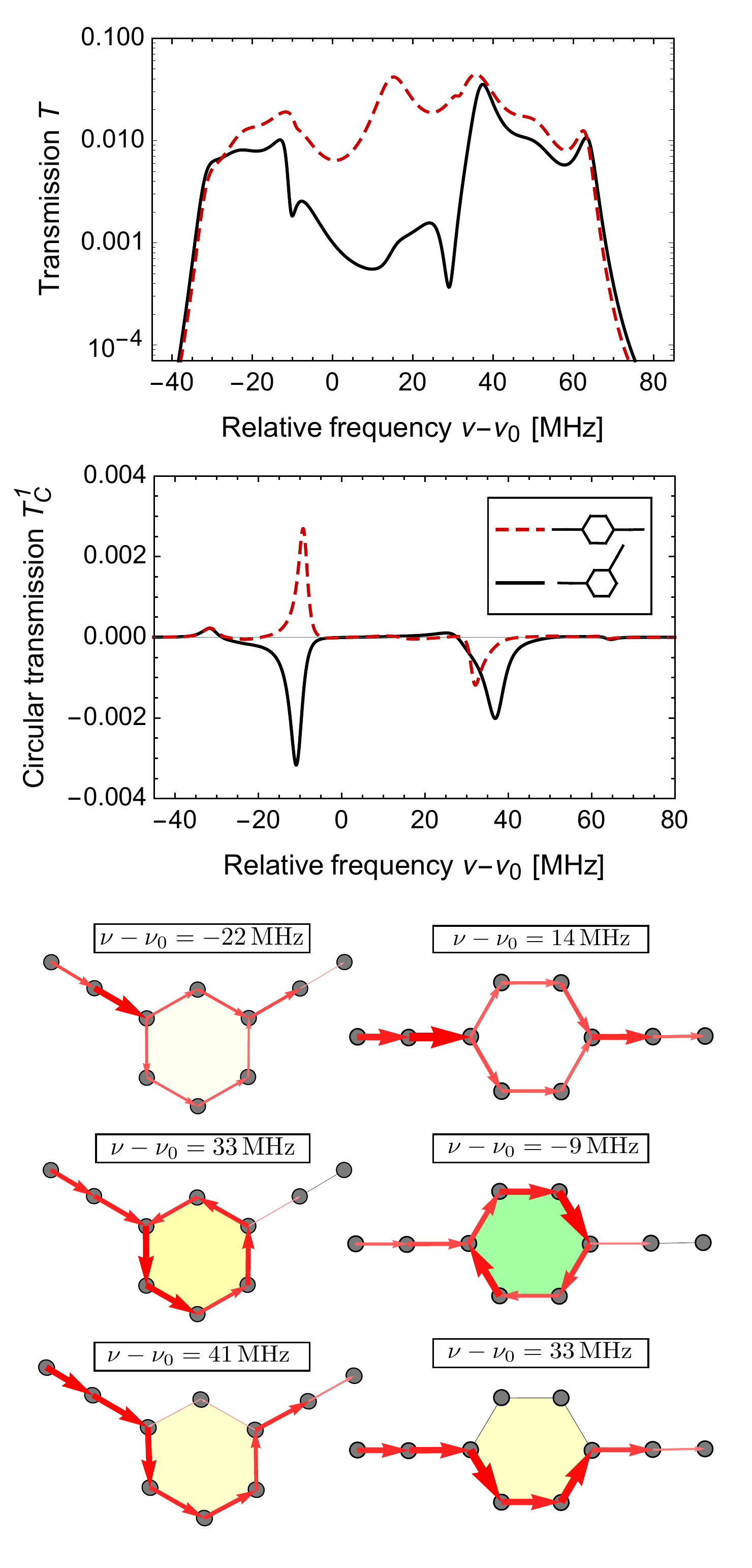}
  \caption{Current flow in the microwave emulation experiment. Top: Transmission and circular
    transmission as a function of the microwave frequency. Bottom: Normalized transmission pathways
    at different microwave frequencies. The microwave experiment clearly confirms the existence of
    circular currents. The transmission pathways as a function of the microwave frequency can be
    found in the movie
    \href{https://www.fis.unam.mx/~stegmann/current-vortices/Benzene-Experiment.mov}{\texttt{Benzene-Experiment.mov}}.}
  \label{fig:5}
\end{figure}

Now, let us turn to the data from the microwave emulation experiment. \Fig{5} shows the transmission
$T$ and circular transmission $T_C^1$ as a function of the microwave frequency as well as the
(normalized) transmission pathways for various microwave frequencies. The transmission shows several
resonances and anti-resonances, similar to the DFT-NEGF calculations. However, their number differs
because the microwave experiment emulates only the non-interacting $\pi$ electron system, while in
the DFT-NEGF calculations all electrons are taken into account. The emulation experiment shows a
considerable reduction of the transmission (in the range from $-10$ to $30 \un{MHz}$) in the meta
configuration, which resembles the band gap observed in the DFT-NEGF calculations. In the para
configuration, the transmission is rather constant while the DFT-NEGF calculations show a (though
less pronounced) band gap. This discrepancy can be explained again by the fact that only a part of
the electronic structure is emulated by the experiment, because the simple tight-binding model
discussed below shows similar trends as the emulation experiment, see \fig{7}. Moreover, the overall
transmission is lower and its features are much less pronounced than in the calculations, which can
be explained by the overall absorption of the microwaves \cite{Stegmann2017, Stegmann2017b}. Note
that the transmission is measured between the antenna at the left end, see \fig{2}, and the antenna
on top of the next-to-last site at the right end. If the antenna is placed on top of the last site
at the right end, the transmission that is passing through the absorber is measured. It follows the
same trends as the transmission to the next-to last site but is damped strongly (approximately one
order of magnitude).

The transmission pathways in the meta-benzene analogue show strong circular currents at certain
frequencies and agree qualitatively with the patterns obtained in the DFT-NEGF study, although a
sign reversal of the circular transmission is not found. In para-benzene, we find transmission
pathways that are similar to the theoretical predictions but surprisingly, we also observe
asymmetric patterns as well as current vortices. Note that in both cases the local transmissions are
not conserved due to the absorption of the microwaves. In particular, the transmissions towards the
drain at the right end decay strongly, because of the absorber on top of the last resonator, which
damps the current that is passing to the measuring antenna (on top of this resonator).

% EXPLANATION OF THE CIRCULAR CURRENTS

\begin{figure}[t]
    \centering
    \includegraphics[scale=0.42]{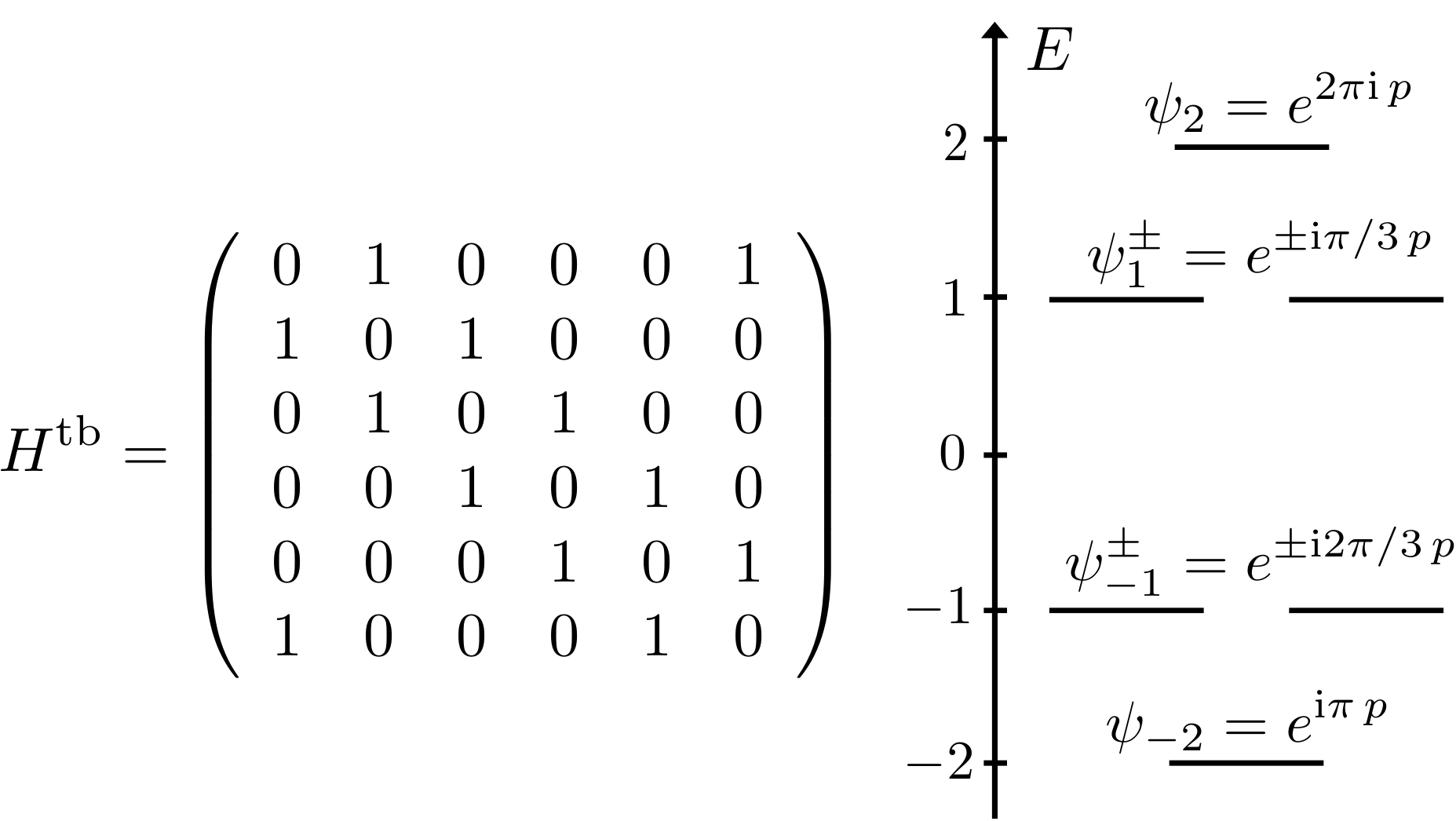}
    \caption{Left: nearest-neighbor tight-binding model of benzene. Right: eigenenergies and
      eigenstates. The variable $p\in (0,1,2,3,4,5)$ numerates cyclically the six carbon atoms and
      generates the six componentes of the eigenstates.}
  \label{fig:6}
\end{figure}

In order to better understand the observed effects, let us consider a simple nearest-neighbor
tight-binding model for the benzene molecule. The Hamiltonian, its eigenenergies and eigenstates are
shown in \fig{6}. By means of these eigenstates, we can calculate the current that is flowing in the
isolated molecule (i.e. in absence of the leads) at the corresponding eigenenergies. The
non-degenerate states at $E= \pm 2$ do not cause any current flow, because they are entirely
real. The degenerate states at $E= \pm 1$ are complex valued and generate non-zero ring currents
with opposite direction of rotation. The same observation holds for the angular momentum%
\footnote{The expected value of the angular momentum of a tight-binding state $\psi$ is determined
  by
  \begin{equation*}
    \braket{\psi | L | \psi} \sim \sum_{i<j} \Im{H^*_{ij}\, \psi_i^* \psi_j}
    \lr{(\v r_i {-}\v r_0) \times (\v r_j {-}\v r_0)},
  \end{equation*} 
  where $\v r_i$ are the positions of the sites (or atoms) with respect to the center $\v r_0$. The
  angular momentum is proportional to the circular current in \eq{6} because
  $G^n_{ij} \sim \psi_i^* \psi_j$.} %
of these states as it is proportional to the circular current in \eq{6} \cite{Rai2010,
  Dhakal2019}. However, note that one could choose the degenerate eigenstates also in such a way
that their ring currents and angular momenta are zero.

\begin{figure}[t]
  \centering
  \includegraphics[scale=0.5]{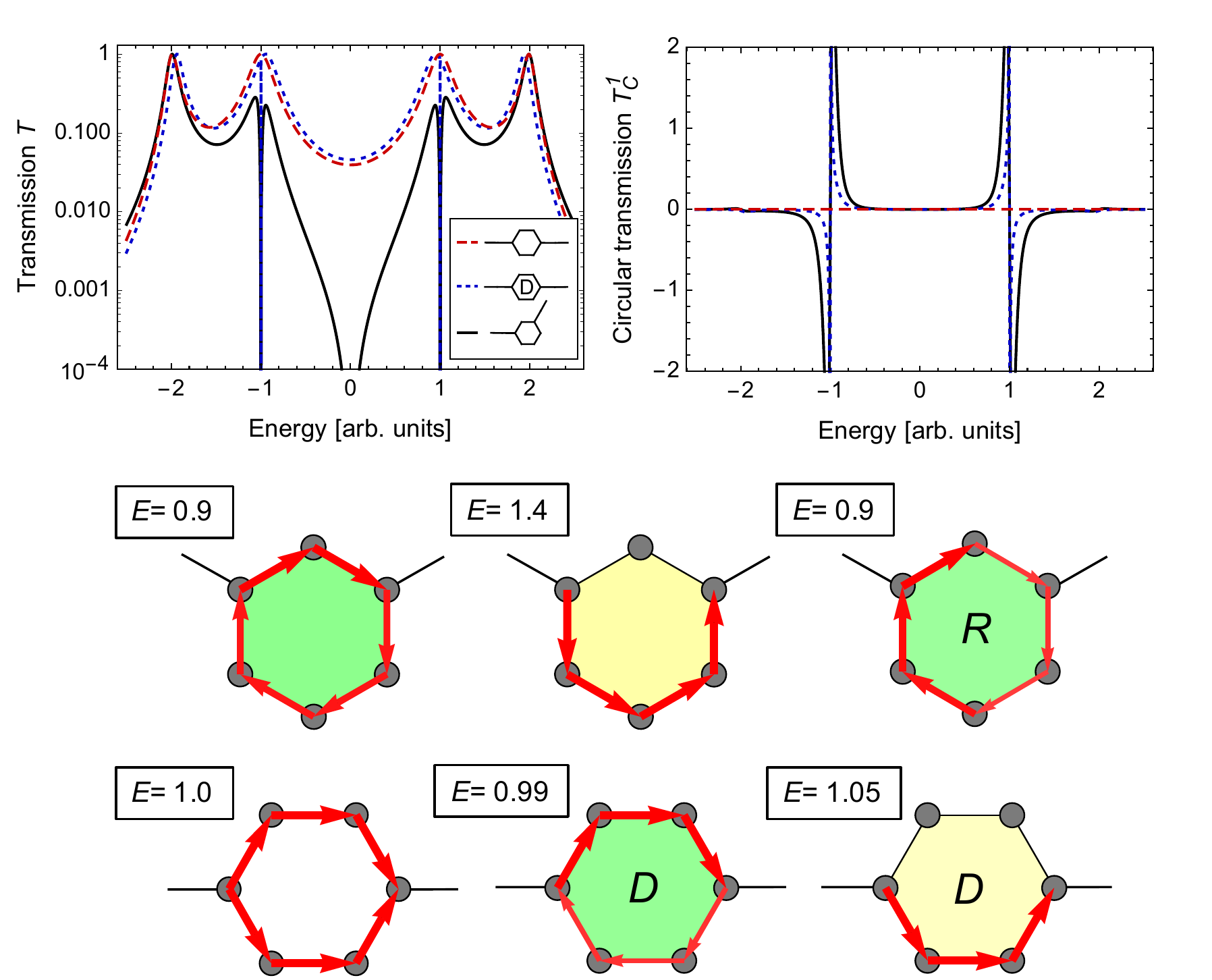}
  \caption{Current flow in the benzene molecule modelled by a simple tight-binding Hamiltonian
    (\fig{6}) and wide-band leads (\eq{7}). The transmission and circular transmission (top) show
    similar features as the microwave emulation experiment. The transmission pathways (bottom) show
    vortices that are obtained also in the DFT-NEGF calculations and the experiment. The pathway
    with the label $R$ has been obtained by taking into account only the two eigenstates close to
    $E=1$. For the pathways with the label $D$, we have introduced random fluctuations ($5\%$) in
    the coupling energies, which induces circular currents. The full dependence of the transmission
    pathways as a function of the electron energy can be found in the movie
    \href{https://www.fis.unam.mx/~stegmann/current-vortices/Benzene-TB.mov}{\texttt{Benzene-TB.mov}}.}
  \label{fig:7}
\end{figure}

The coupling of the molecule to the leads is modelled within the wide-band approximation with the
self-energies
\begin{equation}
  \label{eq:7}
  \Sg_{ij}^{S/D}= -\I \eta \, \dl_{i,j}\,\dl_{i,s/d},
\end{equation}
where $s=1$ is the position of the source, while $d=4$ and $d=3$ give the position of the drain in
the para and meta configuration, respectively. The parameter $\eta=0.2$ controls the coupling
strength of the leads and $\dl_{ij}$ is the Kronecker delta function. This non-Hermitian
perturbation of the system lifts on the one hand the degeneracies of the eigenstates, but on the
other hand all of these eigenstates have zero angular momentum. However, the transmission pathways
calculated by means of \eq{3} agree qualitatively with the DFT-NEGF calculations and the microwave
experiment and confirm the existence of circular currents, see \fig{7}. An analytic study of the
benzene molecular junction that focusses on quantum interference can be found in
Ref.~\cite{Hansen2009}.

To understand the existence of these vortices, we expand \eq{3} in terms of the eigenstates
$\psi_\ap$ of the open system $H+\Sg_S+\Sg_D$,
\begin{equation}
  \label{eq:8}
  T_{ij}(E)= \sum_{\ap,\bt}\text{Im} \biggl[ H_{ij}^*\,
  \frac{\psi_{\ap,s}\, \psi_{\ap,i}\, \psi^*_{\bt,j}\, \psi^*_{\bt,s}}{(E-\eps_\ap) (E-\eps^*_\bt)}\biggr],
\end{equation}
where $s=1$ is the position of the source. Let us start with the meta configuration. Taking into
account in the sum in \eq{8} only the two eigenstates close to $E=\pm 1$, which are degenerate in
absence of the leads, we observe qualitatively the same circular transmission pathways, see in
\fig{7} the benzene molecule marked with $R$. Taking into account in \eq{8} only the diagonal terms
(i.e. $\ap=\bt$), or considering only one of the two eigenstates, gives qualitatively different (and
hence incorrect) transmission pathways without any ring currents. In para benzene, where ring
currents are not possible due to the symmetry of the molecular junction (see above), one of the
degenerate eigenstates at $E= \pm 1$ (i.e. the state
$\psi'_{\pm 1}= \psi_{\pm 1}^+ - \psi_{\pm 1}^-$) is unchanged by the leads, because it has zero
value on the carbon atoms where the leads are attached. This eigenstate does not couple to the leads
and does not contribute to the current flow. The current around $E= \pm 1 $ is determined by a
single eigenstate and does not show any vortices. The reader maybe intereded in a discussion of
related problems using molecular orbitals in Refs. \cite{Jhan2017}.

The fact that the microwave experiment shows circular currents also in the para configuration of the
benzene molecule can be explained by the imperfections of the experiment. The resonance frequency of
the dielectric resonators ($\pm 5 \un{MHz}$) as well as their position ($\pm 0.2 \un{mm}$) vary
slightly due to manufacturing imperfections and limited precision in placing them, respectively. We
take into account these imperfections in the tight-binding calculations by changing randomly the
nearest-neighbor coupling energies by 5\%. This induces circular currents also in para benzene, see
the transmission pathways with the label $D$ in \fig{7}. In terms of \eq{8}, the random
perturbations couple the (previously uncoupled) eigenstates $\psi'_{\pm 1}$ to the leads and hence,
induces current vortices. As the perturbations are small, the coupling is weak and ring currents are
observed only in a narrow window around $E= \pm 1$. Taking into account this type of disorder in the
meta configuration does not change qualitatively the observed current patterns, because a priori all
eigenstates are coupled to the leads.

% ANTHRACENE
\subsection{Anthracene \& naphthalene}

\begin{figure}[t]
  \centering
  \includegraphics[scale=0.5]{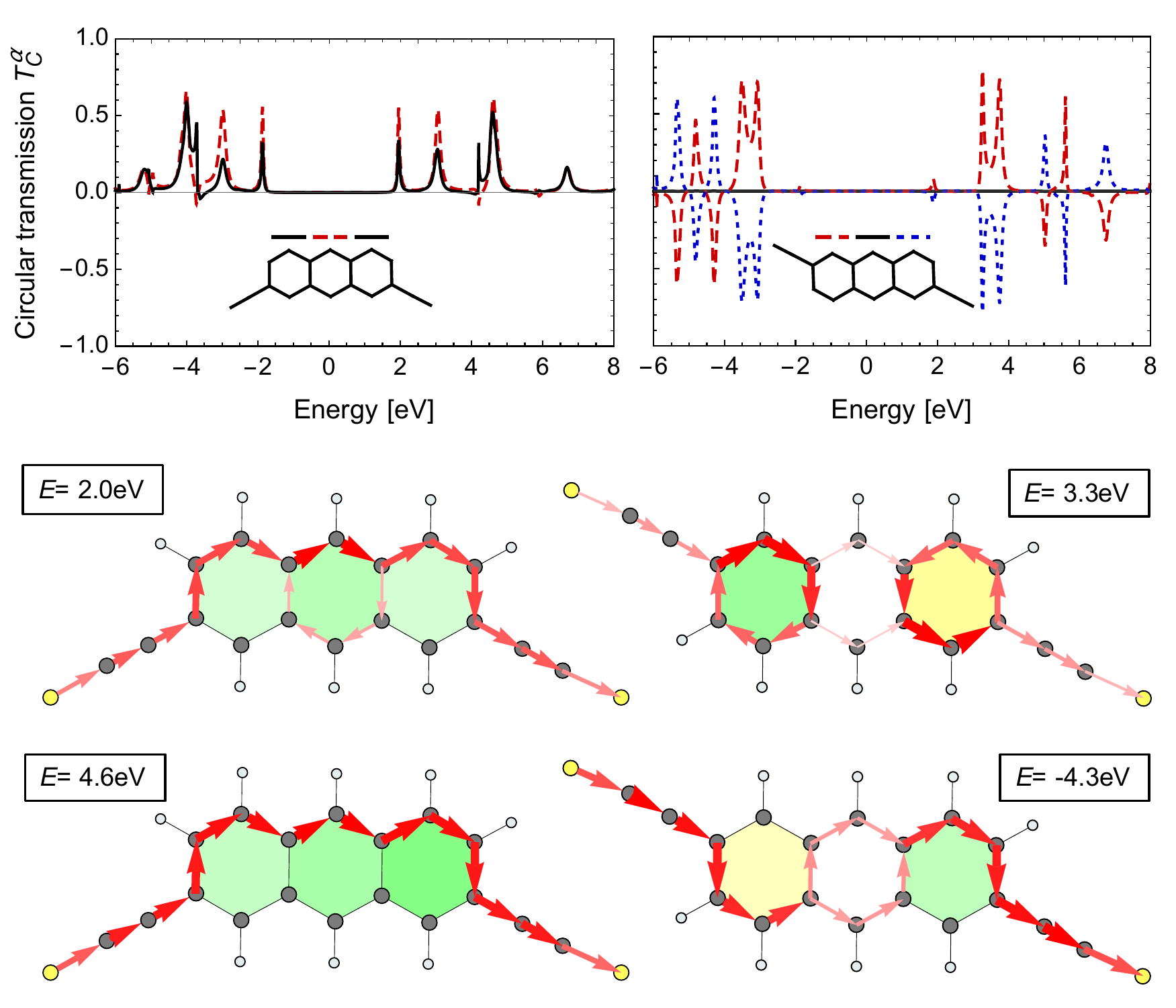}
  \caption{Current flow in the anthracene molecular junction calculated by means of the DFT-NEGF
    method. The meta configuration is shown in the left column, the para configuration in the right
    column. Top: Circular transmission in the three ($\ap \in (1,2,3)$) carbon rings. Bottom:
    Transmission pathways at various electron energies.}
  \label{fig:8}
\end{figure}

\begin{figure}[t]
  \centering
  \includegraphics[scale=0.52]{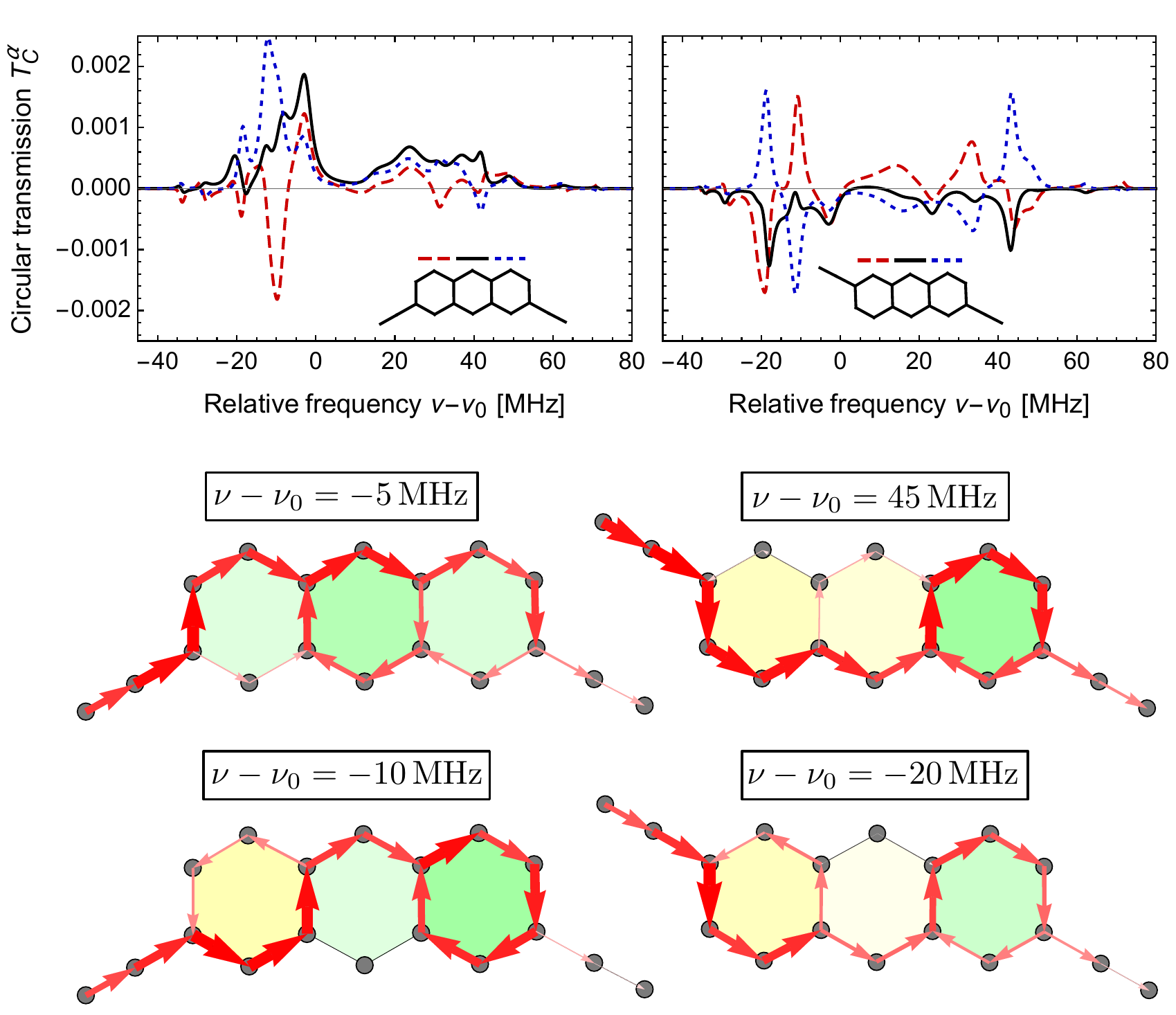}
  \caption{Circular transmission (top) and transmission pathways (bottom) obtained by the microwave
    emulation experiment of the anthracene molecule. Current vortices are observed in both
    configurations of the leads. }
  \label{fig:9}
\end{figure}

Now, let us proceed with longer carbon ring molecules, namely anthracene and naphthalene. The
circular transmission and the transmission pathways, calculated by the DFT-NEGF method, are shown in
\fig{8} for the meta and para configuration of the anthracene molecular junction. The corresponding
results from the microwave emulation experiment are shown in \fig{9}. Theory and experiment confirm
both the existence of circular currents for the two configurations of the leads.

\begin{figure}[t]
  \centering
  \includegraphics[scale=0.52]{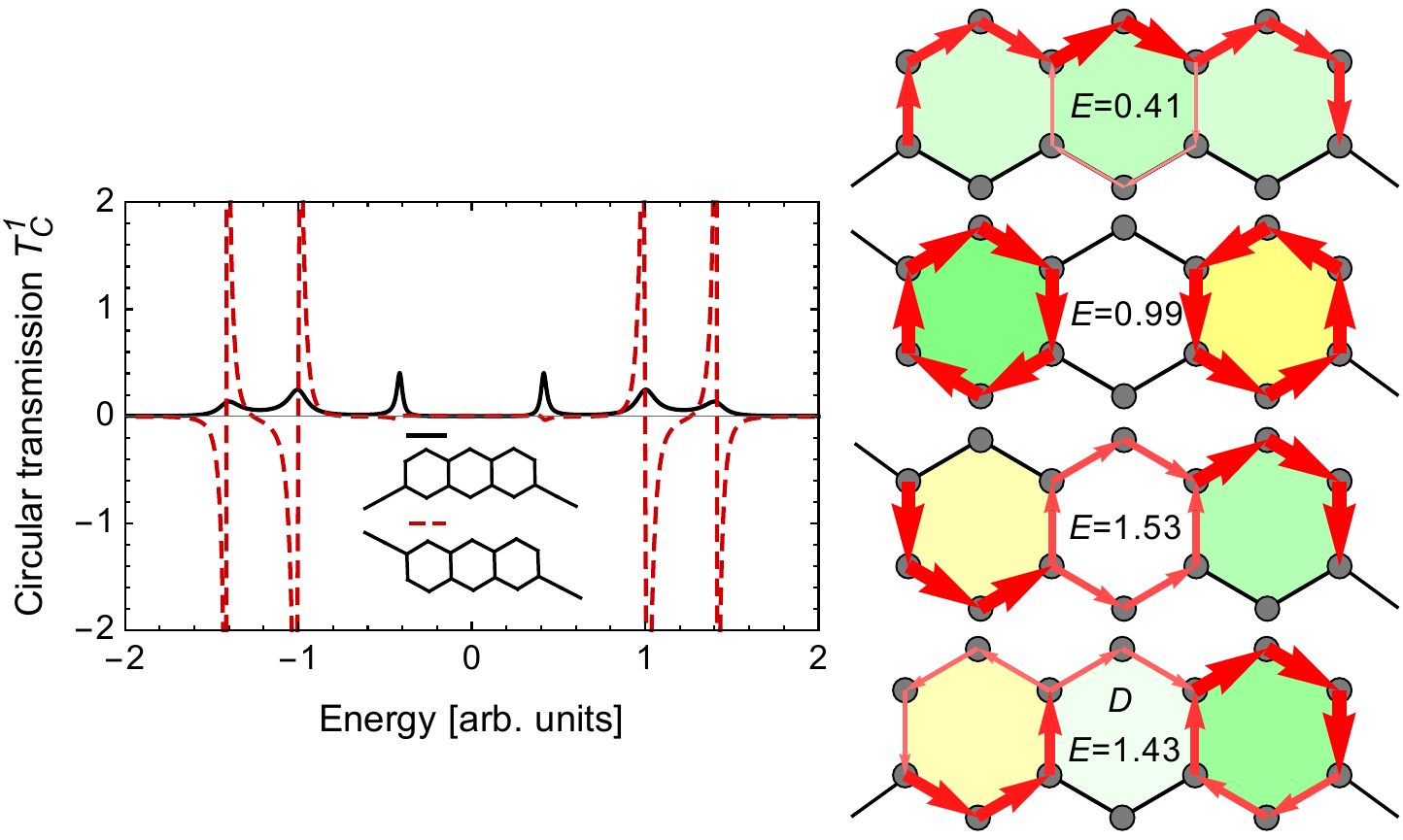}
  \caption{Circular transmission (left) and transmission pathways (right) in the anthracene
    molecular junction, calculated within a simple nearest-neighbor tight-binding model. The
    transmission pathways agree qualitatively with the DFT-NEGF calculations and the microwave
    experiment. The asymmetric transmission pathway labeled with $D$ is obtained by changing randomly
    the coupling energies by 5\%.}
  \label{fig:10}
\end{figure}

The circular currents can be understood by a simple tight-binding Hamiltonian, similar to our
discussion of the benzene molecule, see \fig{6}. In anthracene, we have four two-fold degenerate
eigenstates at the energies $E= \pm 1$ and $E= \pm \sqrt{2}$, where strong current vortices can
appear, see \fig{10}. Let us consider a pair of these degenerate states. When the leads are attached
to the molecule, the degeneracy is lifted but only in the para configuration both states couple to
the leads. Taking into account in \eq{8} only these two states shows that they determine the
transmission pathways in the proximity of their eigenenergy. In particular, the correlations between
the states, taken into account in \eq{8} by the elements with $\ap \neq \bt$, contribute
significantly to the current flow. When the leads are attached in the meta position, one of the
previously degenerate states has zeros at the positions of source and drain and hence, does not
couple to them. The fact that in meta anthracene circular transmission pathways are found, can be
explained by the interplay of states that are non-degenerate in the isolated molecule but nearby in
energy. This also explains why the circular transmissions is much less pronounced in this
configuration of the leads.

In the calculation, the local transmissions and the circular transmission in the three carbon rings
show certain symmetries. For example, in the meta configuration the circular transmission in the
first and third carbon ring are equal $T_C^1= T_C^3$, while in the para configuration they have
opposite sign $T_C^1= - T_C^3$, and zero value in the central ring $T_C^2= 0$. These symmetries
cannot be observed in the microwave experiment due to the different modelling of source and drain
and the imperfections of the experiment (see above). Changing in the tight-binding model randomly
the coupling energies by 5\%, we obtain asymmetric current patterns very similar to those obtained
in the emulation experiment, see for example in \fig{10} the transmission pathway labeled with $D$.

% NAPHTHALENE

Finally, we address briefly the naphthalene molecule, using a simple tight binding model. All
eigenstates of this molecule are non-degenerate and real. Hence, they cannot generate any ring
current nor have any angular momentum. However, when leads are coupled to the molecule and transport
is studied, circular transmission pathways are observed but their shape differs from the pathways
found in benzene and anthracene. In general, they are a consequence of the interplay of all
eigenstates of the open system, see \eq{8}, where the main contributions come from the eigenstates
that are nearby the considered energy.

\begin{figure}[t]
  \centering
  \vspace{9mm}
  \includegraphics[scale=0.5]{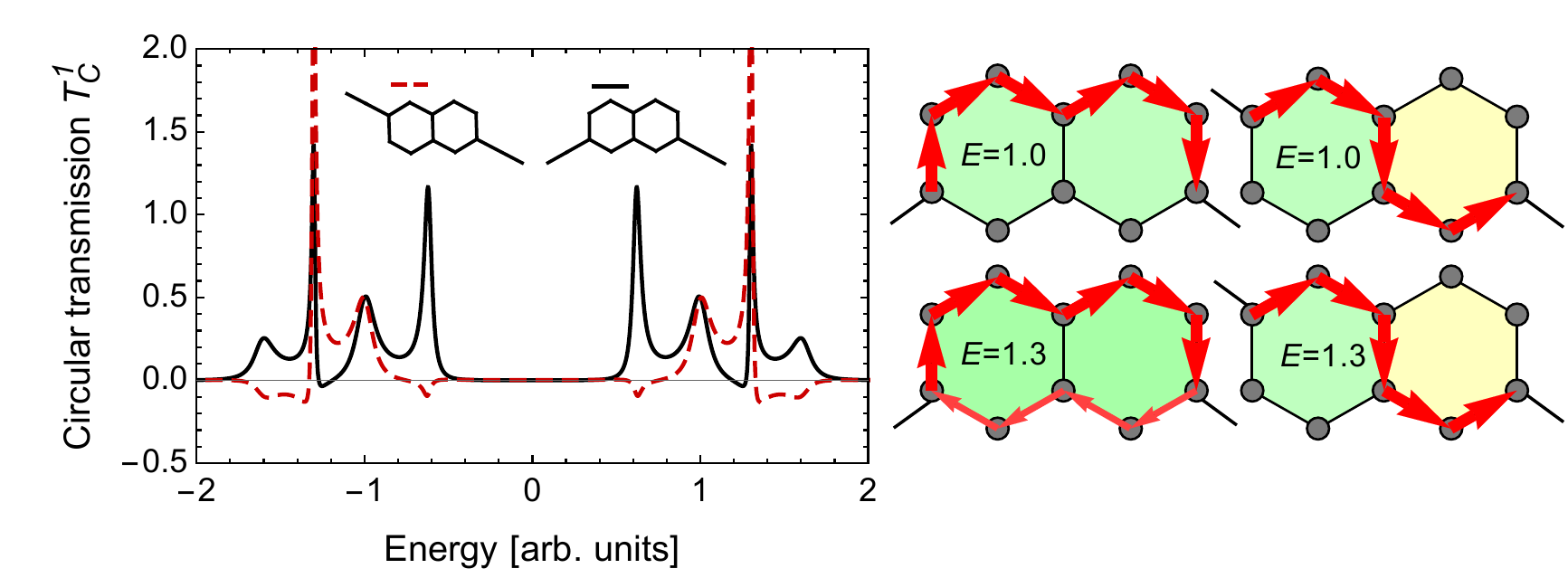}
  \caption{Circular transmission (left) and transmission pathways (right) in the naphthalene
    molecular junction, calculated within a simple nearest-neighbor tight-binding model. Although all
    eigenstates of the isolated molecule are non-degenerate, the open system shows circular
    currents.}
  \label{fig:11}
\end{figure}

\section{Conclusions \& Outlook}
\label{sec:Conclusions}

We have presented an analysis of the local current flow in benzene and anthracene with some remarks
on naphthalene. We have performed state-of-the-art DFT-NEGF calculations of the current flow as well
as microwave emulation experiments that model the non-interacting $\pi$ electron system of the
molecule. Both show qualitatively similar transmission pathways and confirm the existence of current
vortices in certain regions of electron energies, see for example Figures \ref{fig:3} -
\ref{fig:5}. These emulation experiments are -- to the best of our knowledge -- the first direct
measurement of circular currents in ring structures. The circular currents can be understood in
terms of a simple nearest-neighbor tight-binding H\"uckel model, which also connects more deeply the
DFT-NEGF calculations and the emulations experiments, as it approximates the former and describes
qualitatively the latter. Using the spectral decomposition of the local transmissions, see \eq{8},
we have shown that the circular currents can be understood by the interplay of the complex
eigenstates with energies close-by the considered electron energy. In particular, the cross terms
($\ap \neq \beta$) have a significant effect. Degnerate states, as they appear in benzene and
anthracene, generate strong circular currents if both states couple to the leads (like in
meta-benzene). One of these states alone cannot induce a current vortex. We have also shown that
small imperfections and perturbations can couple previously uncoupled states to the leads and hence,
induce current vortices in the system even if symmetry conditions forbid at first their occurrence,
see \fig{7}.

As to the significance of the results we can only speculate, but practical use may result from the
magnetic field caused by vortices of strong circular transmission. These vortices appear close to
transmission maxima and tend to show currents circulating around the complete ring, which imply
stronger and more homogeneous magnetic fields within the ring than currents flowing only on one side
of the ring. A main challenge to detect this magnetic field is not only to realize experimentally
the molecular junctions, but also to shift the Fermi energy close to one of the resonances of the
circular transmission \cite{Liu2016}. We plan to investigate if this can be achieved by designing
specific aromatic molecules, for example by adding substituents \cite{Herrmann2010a} or by changing
the molecular structure \cite{Schlicke2014}. We also expect that current vortices will appear in
larger aromatic carbon molecules, like graphene nano-ribbons, which may be easier to implement in
the experiment.

\section*{Conflicts of interest}
There are no conflicts of interest to declare.

\section*{Acknowledgments}
We gratefully acknowledge funding from CONACYT Proyecto Fronteras 952, UNAM-PAPIIT IA101618,
UNAM-PAPIIT AG100819, DFG project HE \mbox{5675/5-1} and EU project NEMF21. YPO is gratefull for an
UNAM-DGAPA postdoctoral fellowship. We acknowledge use of the MIZTLI super computing facility of
DGTIC-UNAM. We thank Reyes Garcia for computer technical support and Thijs Stuyver for his help with
the Artaios code.

\section*{Electronic supplementary material}
The transmission pathways in the benzene molecular junction as a function of the electron energy (or
microwave frequency) can be found in
\href{https://www.fis.unam.mx/~stegmann/current-vortices/Benzene-DFT.mov}{\texttt{Benzene-DFT.mov}}
for the DFT-NEGF calculations, in
\href{https://www.fis.unam.mx/~stegmann/current-vortices/Benzene-Experiment.mov}{\texttt{Benzene-Experiment.mov}}
for the microwave experiment and in
\href{https://www.fis.unam.mx/~stegmann/current-vortices/Benzene-TB.mov}{\texttt{Benzene-TB.mov}}
for the tight-binding calculations.

\bibliography{edcm}

\end{document}